\documentstyle[epsfig]{article}
\def\cl{\centerline} \def\ni{\noindent} 
\def\vs{\vskip} \def\hs{\hskip}  \def\r#1{$^{[#1]}$}

\def\beq{\begin{equation}}   \def\eeq{\end{equation}}
\def\EE{e$^+$e$^-$}

\begin{document}
\null{}\vs-2.5cm

\hs11cm{HZPP-9809}

\hs11cm{Dec. 10, 1998}
\vs1cm

\begin{center}
{\Large\bf LEVY  STABILITY  INDEX
\vskip0.5cm

FROM  MULTIFRACTAL  SPECTRUM
\footnote{This work is supported in part by the National 
Natural Science Foundation of China.  (NSFC) under Grant No.19575021. }}

\vskip1.2cm

{\large Hu Yuan \ \ \ \ \ \  Yu Meiling \ \ \ \ \ \ \ Liu Lianshou}

{\small Institute of Particle Physics, Huazhong Normal University, 
Wuhan 430079 China}

{\small Tel: 027 87673313 \qquad FAX: 027 87662646 
\qquad email: liuls@iopp.ccnu.edu.cn}

\end{center}

\vskip3.2cm
\begin{center}
\begin{minipage}{125mm}
\vskip 0.5in
\begin{center}{\Large ABSTRACT}\end{center}
{\hskip0.6cm  
A method for extracting the Levy
stability index $\mu$ from the multi-fractal spectrum $f(\alpha)$
in high energy multiparticle production is proposed. This index is an
important parameter, characterizing the non-linear behaviour of 
dynamical fluctuations in high energy collisions.  
Using the random cascading $\alpha$ model as example,
the validity of this method is tested. It is shown that this method,
basing on a linear fit, is consistent with and more accurate than
the usual method of fitting the ratio of $q$th to 2nd 
order multi-fractal (R\'enyi) dimensions to the Peschanski formula.}
\end{minipage}
\end{center}
\vskip 2cm
\ni
\vskip0.5cm

\ni

\hskip0.5cm 
\ \ R\'enyi dimension
\newpage

Recently, the search, starting from eighties\r{1}, for the non-linear phenomena 
---  intermittency or fractality in the multiparticle final states of high 
energy collisions\r{2} has had breakthrough\r{3}.  It is found that
dynamical fluctuations do exist in high energy hadron-hadron collisions, 
but is aniosotropic rather than isotropic\r{4}.  Power law in the 
higher-dimensional factorial moment versus partition number has been 
observed in hadron-hadron collisions, when the anisotropy of dynamical 
fluctuations is taken into account properly\r{5,6}. 
On the other hand, the dynamical fluctuations in 
\EE collisions are found to be approximately isotropic\r{7}. 

\def\ss{\hskip2pt}
As a (self-affine or self-similar) fractal system, the multiparticle final 
state in high energy collisions can be characterised by an important 
parameter --- the Levy stability index $\mu$\r{8}. This parameter 
tells us the behaviour of elementary fluctuations at the tail of distribution.  
To extract its value from exprimental data more reliably 
is an essential task for the understanding of fluctuation dynamics.

Previously, people usually use the anomalous scaling of factorial moment\r{9}
\beq  
F_q=\frac{1}{M} \sum \limits_{m=1} ^M \frac {\langle  n_m (n_m-1) 
\cdots (n_m-q+1) \rangle} {\langle n_m \rangle ^q} 
\eeq
to get the scaling index (IM index) $\phi_q$ and multi-fractal (R\'enyi) 
dimensions\r{10} 
\beq  
 D_q = 1- \frac{\phi_q}{q-1}
\eeq
of integer order $q$, and then fit the ratio $(1-D_q)/(1-D_2)$ to the 
formula\r{11}
\beq   
\frac{1-D_q}{1-D_2}=\frac{1}{q-1} \frac{q^\mu -q}{2^\mu -2}  
\eeq
to get the value of $\mu$.
In practice only a few indices $\phi_q$ of integer $q>2$ ($q=3,4,5$ e.g.) 
can be obtained and the resulting value of $\mu$ basing on a fit of very
few points to the complicated formula (3) is obviously unsatisfactory.

In order to establish an alternative way for extracting the value of $\mu$,
let us notice that 
the $q$th order multi-fractal R\'enyi dimension $D_q$ of 
any order $q$ can be obtained from the multifractal spectrum $f(\alpha)$  
through\r{10}
\beq   
D_q = \frac{1}{q-1} (q\alpha  - f(\alpha)).
\eeq
Therefore, it should be able to get the Levy stability index $\mu$ 
directly from the multifractal spectrum $f(\alpha)$. This short note 
is aimed to discuss this problem.

It can be seen from Eq.(4) that
\beq  
 \tau_q =(q-1)D_q 
\eeq
is the Legandre transform of $f(\alpha)$. Therefore, we have\r{12}
\beq  
 \alpha_q=\frac{d\tau_q}{dq} , 
\eeq
\beq  
f(\alpha)=q\alpha_q-\tau_q .
\eeq

From Eq.(3) we have
$$ D_q= 1-A \frac{q^\mu-q}{q-1} , $$
where $ A = (1-D_2)/(2^\mu -2) . $
Inserting into Eq.(5) and making use of Eq's. (6),(7) we get
\beq 
 \tau_q        
   = -1+(1+A)q-Aq^\mu  ,
\eeq
\beq 
\alpha_q=1+A-\mu Aq^{\mu-1} ,
\eeq
\beq 
 f(\alpha)=  
1-(\mu-1)Aq^\mu .
\eeq
Therefore, defining
\beq 
B=1+A=1+\frac{1-D_2}{2^\mu -2} , 
\eeq
we finally get
\beq  
1-f(\alpha) \propto (B-\alpha)^{\mu/(\mu-1)} \qquad 
\mbox{\rm when} \ \alpha < B .
\eeq

Thus we propose the following procedure for getting the Levy stability index 
$\mu$ from the multi-fractal spectrum $f(\alpha)$ :

(1) Find out the value of $\alpha$ where $f(\alpha)=1$, call it $B$ .

(2) Fit the $\alpha<B$ part of ln$(1-f(\alpha))$ versus ln($B-\alpha)$ 
to a straight line and get the slope $C$. 

(3) Get Levy stability index as 
\beq  
 \mu = \frac{C}{C-1} .
\eeq

As an example, we apply this method to the random cascading 
$\alpha$-model\r{13}.  This model describes each multiparticle 
event as a series of steps, in which the initial phase space region 
$\Delta$ is repeatedly divided into $\lambda=2$ parts.        
After $\nu$ steps we get $M= {2^ \nu}$ sub-cells of size
$\delta = {\Delta}/{M}$.
At each step $s$ the normalized particle density is obtained in
each of the two parts by multiplication of the normalized density
in the step $s-1$ by a particular value of the random variable
${\omega}_{\nu{ j_ \nu}}$, where
$j_ \nu$ is the position of a sub-cell at the ${\nu}$th step
(1$ \leq$ $j_{\nu}$ $\leq$ ${2^\nu}$).
The elementary fluctuation probability $\omega$ is chosen as\r{13}
\beq  
{\omega}_{\nu,2j-1}={\frac{1}{2}}(1+{\alpha}r) \ \ \ \  ;  \ \ \ \
{\omega}_{\nu,2j}={\frac{1}{2}}(1-{\alpha}r), 
\eeq
in which, $r$ is a random number distributed uniformly in the interval
$[-1,1]$, $\alpha$ is a model parameter, characterizing the strength
of fluctuations. In our calculation, we have chosen its value as
$\alpha=0.3$.

The resulting multi-fractal spectrum $f(\alpha)$ in this model are shown 
in Fig.~1. The value of $\alpha$ for $f(\alpha)=1$ is found to be $B=1.0113$.
The fitting of $(1-D_q)/(1-D_2)$ to Eq.(3) and the linear fit of
ln$(1-f(\alpha))$ versus ln($B-\alpha)$ are shown in Fig's.2 and 3 respectively.
The resulting Levy indices $\mu$ from these two methods with different regions
of $q$ together with the corresponding $\chi^2$/DF are 

\begin{tabbing}
AA \= AAAAAAA\= AAAAAAAAAAA \= AAAAAAAAAAAA \= \kill
Fit from $(1-D_q)/(1-D_2)$ \\
\>  ($q=2,3,4,5$) \>\> 
$\mu=1.929 \pm 0.059$ \> $\chi^2$/DF = 0.016 /2. \\
Linear fit of ln$(1-f(\alpha))$ versus ln($B-\alpha)$  \\
\>  ($q=0.1$--$5$, \> $\alpha=1.009$--0.909) \> 
$\mu=1.962 \pm 0.013$  \> $\chi^2$/DF = 0.758 /48. \\
\>  ($q=2$--$5$, \> $\alpha=0.967$--0.909) \>  
$\mu=1.950 \pm 0.022$  \> $\chi^2$/DF = 0.107 /29. \\
\>  ($q=3$--$5$, \> $\alpha=0.945$--0.909) \>   
$\mu=1.940 \pm 0.015$  \> $\chi^2$/DF = 0.015 /19. 
\end{tabbing}
\vskip-0.5cm

\ni cf. Fig.4.

Thus we see that the method we proposed for extracting the Levy
stability index $\mu$ from the multi-fractal spectrum $f(\alpha)$, basing
on a linear fit, is consistent with the usual method of fitting the ratio  
of $q$th to 2nd order multi-fractal (R\'enyi) dimensions to the Peschanski 
formula Eq.(3). It is more accurate and reliable and is reasonably stable 
with respect to the region of $q$ or $\alpha$ used in fitting. 

This method can readily be applied to real experimental data analysis 
as a better way for getting Levy stability index. 

\vskip0.5cm

\ni Acknowledgement

The authors thank Fu Jinghua and Liu Feng for helpful discussions.

\vs3cm
\ni
{\Large\bf Figure Captions}

\vskip0.5cm
\ni{{\bf Fig.1} \ Multi-fractal spectrum}

\ni{{\bf Fig.2} \ Fitting of $(1-D_q)/(1-D_2)$ to Eq.(3) for $q=2,3,4,5$}

\ni{{\bf Fig.3} \ Linear fit of ln$(1-f(\alpha))$ vs. ln$(B-\alpha$)
for $q=0.1$ -- $5$}  

\ni{{\bf Fig.4} \ The resulting $\mu$ value from different ways of fitting
 together with the $q$ region used }

\qquad{and the $\chi^2$/DF}

\newpage

\begin{picture} (260,240) 
\put(-140,-230)   
{\epsfig{file=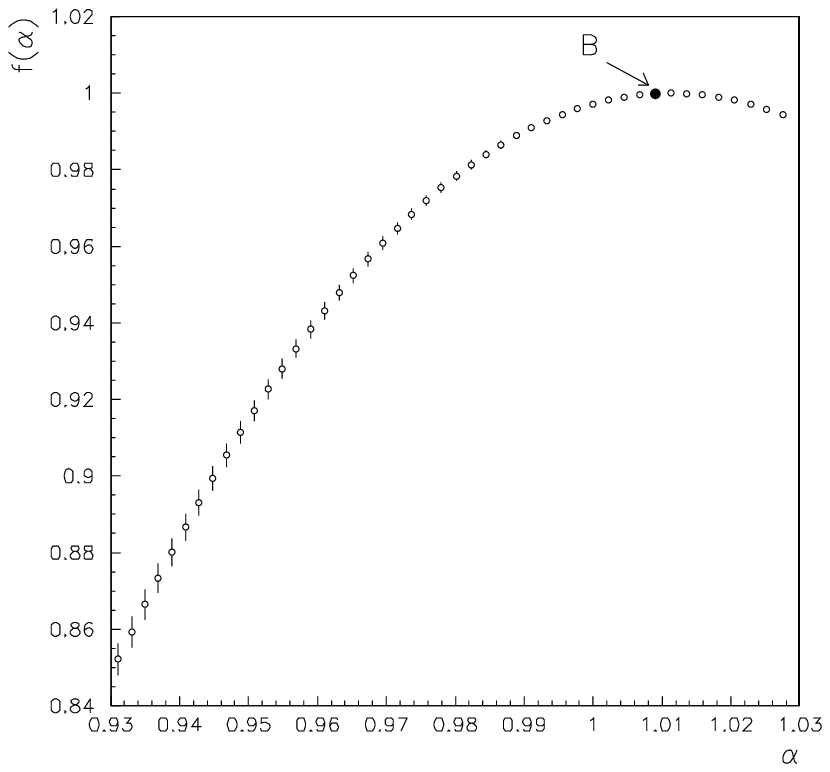,bbllx=0cm,bblly=0cm,
           bburx=8cm,bbury=6cm}}  
\end{picture}
\vs-4.0cm
\vskip1.5cm
\cl{ {\bf Fig.1} \ Multi-fractal spectrum}
\vs-13cm
\begin{picture} (260,240) 
\put(-140,-720)   
{\epsfig{file=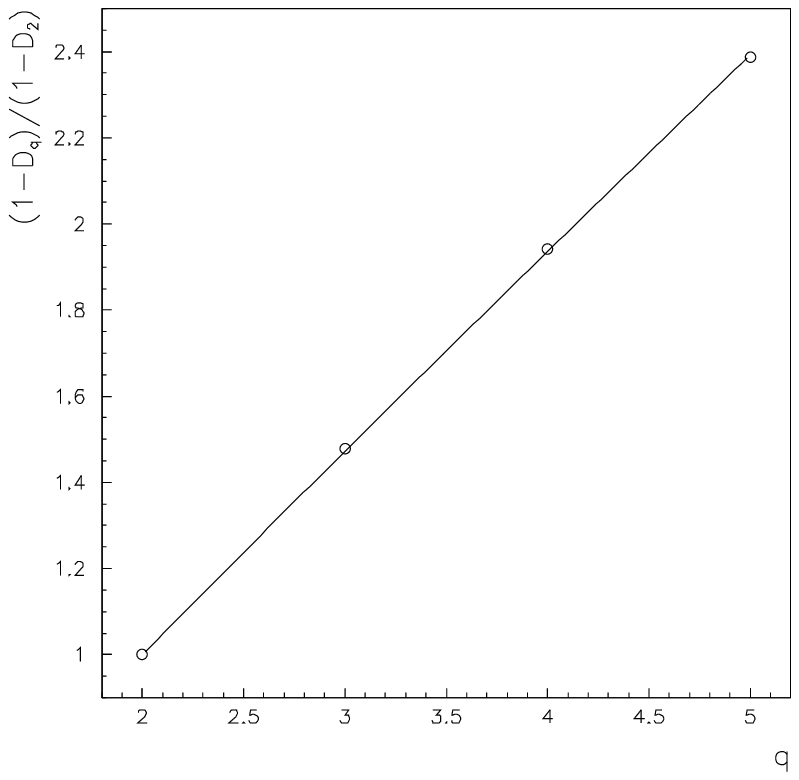,bbllx=0cm,bblly=0cm,
           bburx=8cm,bbury=6cm}}  
\end{picture}

\vs15.0cm
\cl{{\bf Fig.2} \ Fitting of $(1-D_q)/(1-D_2)$ to Eq.(3) for $q=2,3,4,5$}

\newpage

\begin{picture}(260,240)
\put(-155,-230)
{\epsfig{file=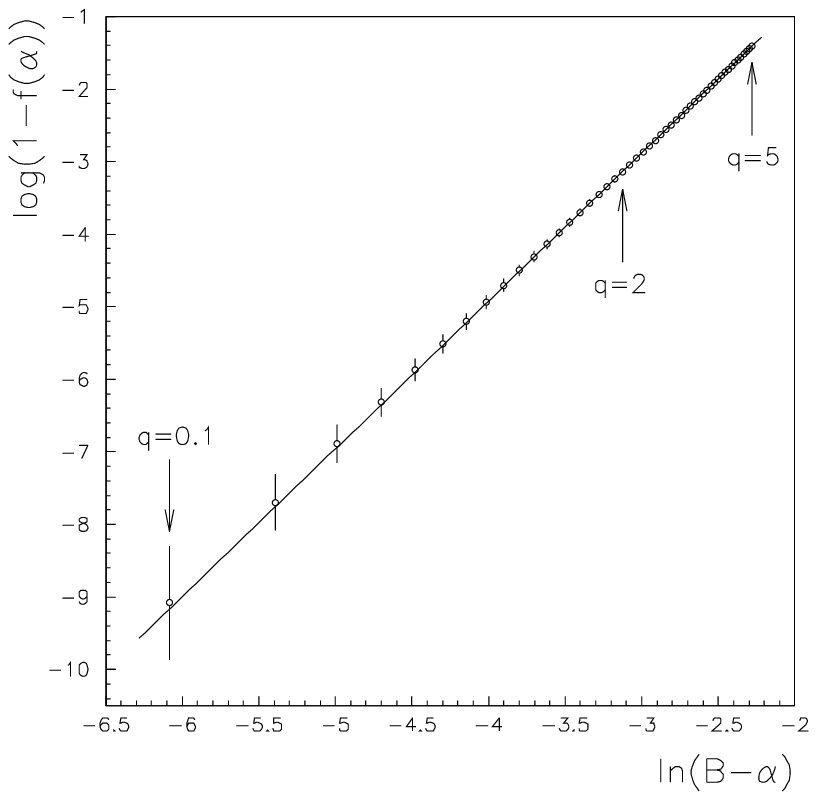,bbllx=0cm,bblly=0cm,
	   bburx=8cm,bbury=6cm}}
\end{picture}

\vs -1.5cm
\cl{ {\bf Fig.3} \ Linear fit of ln$(1-f(\alpha))$ vs. ln$(B-\alpha$)
for $q=0.1$ -- $5$}  

\begin{picture}(260,240)
\put(-155,-280)
{\epsfig{file=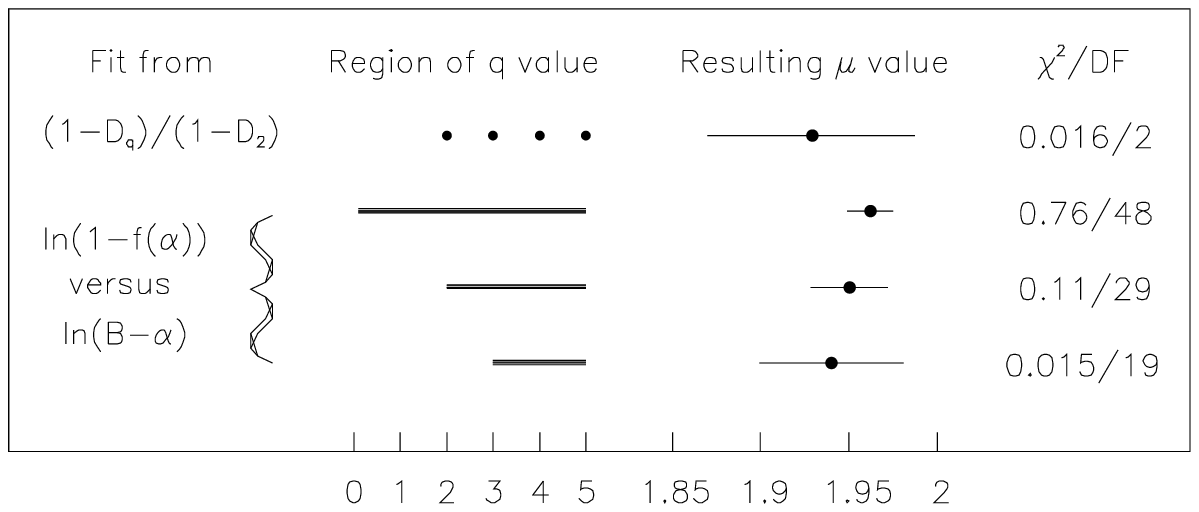,bbllx=0cm,bblly=0cm,
	   bburx=8cm,bbury=6cm}}
\end{picture}

\vs-1.0cm
\cl{{\bf Fig.4} \ The resulting $\mu$ value from different ways of fitting}
\cl{\bf together with the $q$ region used and the $\chi^2$/DF}

(\end{document}